\def\be{\begin{equation}}
\def\ee{\end{equation}}
\def\bea{\begin{eqnarray}}
\def\eea{\end{eqnarray}}

\def\ra{\rangle}

\documentclass[aps,prl,twocolumn,amsmath,amssymb,superscriptaddress]{revtex4}
\usepackage{epsfig,graphicx}
\usepackage{bm}

\begin{document}

%\preprint{draft}
%\draft

\title{Undoing a weak quantum measurement of a solid-state qubit}

\author{Alexander N. Korotkov}
\affiliation{ Department of Electrical Engineering, University of California,
Riverside, CA 92521-0204, USA}
\author{Andrew N. Jordan}
\affiliation{Institute for Quantum Studies, Texas A\&M University, College
Station, TX 77843-4242, USA}

\date{\today}

%\maketitle

\begin{abstract}
We propose an experiment which demonstrates the undoing of a weak continuous
measurement of a solid-state qubit, so that any unknown initial state is
fully restored. The undoing procedure has only a finite probability of
success because of the non-unitary nature of quantum measurement, though it
is accompanied by a clear experimental indication of whether or not the
undoing has been successful. The probability of success decreases with
increasing strength of the measurement, reaching zero for a traditional
projective measurement. Measurement undoing (``quantum un-demolition'') may
be interpreted as a kind of a quantum eraser, in which the information
obtained from the first measurement is erased by the second measurement,
which is an essential part of the undoing procedure. The experiment can be
realized using quantum dot (charge) or superconducting (phase) qubits.
\end{abstract}
%\pacs{PACS numbers: }
%\pacs{73.23.-b; 03.65.Ta; 85.35.-p}

\maketitle
%\narrowtext
%\vspace{1ex}
%\vspace{0.6cm}

A careless scientist accidently turns on a quantum detector, disturbing
a precious, unknown, quantum state.  Dismayed by this event, he desperately
asks if there is a way to get the state back. Is it possible to {\it undo} a quantum measurement?
According to the traditional theory of projective quantum measurement \cite{Neumann},
the answer is no:  wavefunction collapse is irreversible, the original state
is gone forever and is impossible to resuscitate.
%it is generally impossible to restore the premeasured state.
%except in very special cases (for example, if the premeasured state belongs
%to a known set and measurement result corresponds to only one state in this
%set).
However, as will be presently discussed, the situation is different for
weak %incomplete (partial, continuous, weak, POVM, etc.)
quantum measurements \cite{Nielsen,weak-meas,Kor-99,weak-ss}. It is possible
to fully restore any unknown, pre-measured state, though with probability
less than unity. Such undoing of the measurement disturbance [which we will
also refer to as a quantum un-demolition (QUD) measurement] can be
accomplished by making an additional weak measurement, which ``erases'' the
information obtained from the first measurement (somewhat similar to the
quantum eraser of Scully and Dr\"uhl \cite{Scully-82}). The success of the
undoing procedure is indicated by observing a certain result of the second
measurement, in which case the unknown pre-measured state is fully restored.
The probability of successfully undoing the quantum measurement decreases
with increasing strength of the measurement, tending to zero for a
projective measurement.

The possibility of physically undoing (or reversing) a quantum
measurement has been previously discussed by Koashi and Ueda
\cite{Ueda-99}, who have proposed a quantum optics
photon-counting implementation of the idea, using the Kerr effect.
Reversible measurement has also been discussed by others
\cite{Ban-01,D'Ariano-03} (see also the closely related articles
\cite{reversib}), though mainly from a more formal perspective. In
this Letter, we investigate the undoing of {\it continuous} weak
measurements, particularly applied to solid state qubits. We first
consider a quantum double dot qubit, measured by a quantum point
contact, a system of extensive experimental investigation
\cite{DDexp}. For this system,  we discuss how to practically undo
the measurement, and calculate the undoing probability, as well as
the mean undoing time. The second system we consider is a
superconducting ``phase" qubit \cite{Martinis}, measured by a nearby
SQUID. Coherent non-unitary evolution due to measurement has
recently been experimentally verified in this system \cite{Katz}. We
describe the undoing procedure for the phase qubit, and calculate
the undoing probability, obtaining a result similar to the quantum
dot system. The undoing procedure described for the phase qubit is
only slightly more complicated than the experiment already done,
providing a promising candidate for experimental verification. We
briefly discuss the general theory of QUD measurement, and show that our
specific results maximize the general undoing probability, thus
constituting ideal measurement reversal.

\begin{figure}[tb]
  \centering
  \includegraphics[angle=0,width=8.0cm]{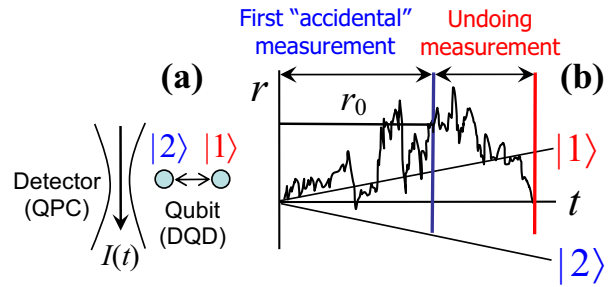}
  \caption{(Color online).
(a) Schematic of a double-quantum-dot (DQD) charge qubit continuously
measured by a quantum point contact (QPC).
 (b) Illustration of the measurement undoing procedure for the charge qubit.
  The slanted lines indicate the detector output in the absence
  of noise, if the qubit is in state $\vert 1\ra$ or $\vert 2 \ra$.
  The initial ``accidental" measurement can be undone by waiting until the
  stochastic measurement result $r(t)$ crosses the origin.
  }
  \label{ok}
\end{figure}

{\it Charge qubit.}--- A double-quantum-dot (DQD) qubit, measured
continuously by a symmetric quantum point contact (QPC) \cite{Gurvitz}
[Fig.~1(a)] has been extensively studied in earlier papers. The measurement
is characterized by the average currents $I_1$ and $I_2$ corresponding to the
qubit states $|1\rangle$ and $|2\rangle$, and by the shot noise spectral
density $S_I$ \cite{S-normalization}. We treat the additive detector shot
noise as a Gaussian, white, stochastic process, and assume the detector is
in the weakly responding regime, $|\Delta I| \ll I_0$, where $\Delta I=
I_1-I_2$ and $I_0=(I_1+I_2)/2$, with QPC voltage bias larger than all other
energy scales,
%(\Delta I)^2/S_I$ (here $\hbar =1$ and $e$ is the electron charge),
so that the measurement process can be described by the
quantum Bayesian formalism \cite{Kor-99}.

    Let us first assume that there is no qubit Hamiltonian evolution
%$\varepsilon=H=0$, so that the qubit state evolves due to the measurement only
(this can also be effectively done using ``kicked'' QND measurements
\cite{Jordan}). As was shown in \cite{Kor-99}, the QPC is an ideal quantum
detector (which does not decohere the measured qubit), so that the evolution
of the qubit density matrix $\rho$ due to continuous measurement preserves
the quantity $\rho_{12}/\sqrt{\rho_{11}\rho_{22}}$, while the diagonal matrix
elements are normalized at all times and evolve according to the classical
Bayes rule:
     \begin{equation}
\frac{\rho_{11}(t)}{\rho_{22}(t)}=\frac{\rho_{11}(0)\exp [-(\bar{I}(t)-I_1)^2
t/S_I]}{\rho_{22}(0)\exp [-(\bar{I}(t)-I_2)^2 t/S_I]} =
\frac{\rho_{11}(0)}{\rho_{22}(0)}\, e^{2r(t)},
    \label{Bayes-DD}
    \end{equation}
where $\bar{I}(t)=[\int_0^t I(t')\, dt']/t$ and we define the {\it
measurement result} as $r(t)=[\bar{I}(t)-I_0]\, t\Delta I/S_I$. For times
much longer than the ``measurement time'' $T_m= 2 S_I/(\Delta I)^2$ (the time
scale required to obtain a signal-to-noise ratio of 1), the average current
$\bar{I}$ tends to either $I_1$ or $I_2$ because the probability density
$P({\bar I})$ of a particular $\bar{I}$ is
    \be
    P({\bar I})=\sum\nolimits_{i=1,2}
\rho_{ii}(0)\sqrt{t/\pi S_I}\, \exp [-({\bar{I}-I_i})^2 t/S_I].
\label{output}
    \ee
Therefore $r(t)$ tends to $\pm \infty$, continuously collapsing the state to
either $|1\rangle$ (for $r\rightarrow \infty$) or $|2\rangle$ (for
$r\rightarrow -\infty$). Critical to what follows, notice that if $r(t)=0$ at
some moment $t$, then the qubit state becomes exactly the same as it was
initially, $\rho (t)=\rho(0)$.  This curious fact corresponds to an equal
likelihood of the states $\vert 1\rangle$ and $\vert 2\rangle$, and
therefore provides no information about the qubit.

  {\it Measurement undoing for the charge qubit.}---
Let the outcome of the ``accidental" first measurement be $r_0$. The
previous ``no information" observation suggests the following strategy:
continue measuring, with the hope that after some time $t$ the stochastic
result of the second measurement $r_u(t)$ becomes equal to $-r_0$, so the
total result $r(t)=r_0+r_u(t)$ is zero, and therefore the initial qubit
state is fully restored. If this happens, the measuring device is
immediately switched off and the undoing procedure is successful [Fig.~1(b)].
However $r(t)$ may never cross the origin, and then the undoing attempt
fails.

The success probability $P_s$ for this procedure may be calculated by
noticing that the nondiagonal matrix elements of $\rho$ do not enter the
probability of the detector output (\ref{output}) (this is true only in the
case of zero or QND-eliminated qubit Hamiltonian), and therefore the
calculation is identical for a classical bit with probabilities
$P_{1,2}=\rho_{11,22}(0)$ of being in state ``1'' or ``2''. These
probabilities should be updated (using the classical Bayes formula) with the
information obtained from result $r_0$: $\tilde{P}_1=P_1
e^{r_0}/[P_1e^{r_0}+P_2 e^{-{r_0}}]$, $\tilde{P}_2=1-\tilde{P}_1$. Assume
for definiteness $r_0>0$.  We will now calculate the probability that the
random variable $r(t)$ crosses the origin at least once, known in stochastic
physics as a first passage process. It follows from (\ref{output}) that both
cases may be described by two different random walks with the initial
condition $r=r_0$ at $t=0$, described by the Green function solution of the
two Fokker-Planck equations
    \be
\partial_t G_i(r,t) = - v_i \, \partial_r G_i + D \, \partial_r^2 G_i + \delta(r-r_0)\,
\delta(t),
\label{G}
    \ee
    supplemented with an absorbing boundary condition at the
origin, where $D=1/(2 T_m)$ is the diffusion coefficient, and
$v_{i}=(-1)^{i+1}/T_m$ are the two different drift velocities, depending on
the bit state. Equations (\ref{G}) may be solved with standard methods
\cite{redner} and from the solution, the first passage time distribution
$P^{(i)}_{\rm fpt}(t)$ is found
from the probability current flux at the origin, %, $P^{(i)}_{\rm fpt}(t)=-v_i G_i+D \partial_x G_i\vert_{x=0}$.
\be P^{(i)}_{\rm fpt}(t) = \frac{r_0}{\sqrt{4 \pi D t^3} } \exp
\left[ - (r_0 + v_i t)^2/(4 D t) \right]. \label{fptd} \ee The
probability that $r=0$ is ever crossed is found by integrating
(\ref{fptd}) over all positive time to obtain $P_{c,1}=\exp (-v_1
r_0/D)=\exp (-2r_0)$ for the crossing probability if $i=1$, and
$P_{c,2}=1$ if $i=2$.  This result is intuitive because starting at
$r_0>0$, a negative drift velocity must cause an eventual crossing,
while a positive drift velocity can only occasionally be beaten by
the noise term.
%{\bf (sentence can be removed in proofs)}
The mean first passage time may also be calculated from (\ref{fptd}) to
obtain $t_{c,i}=r_0/\vert v_i \vert$, averaging only over successful
attempts. Analogous results for $r_0<0$ are straightforward.

Combining these results, the probability
$P_s = {\tilde P}_1 P_{c,1} + {\tilde P}_2 P_{c,2}$ for a successful
quantum undemolition measurement is
\begin{equation}
P_s = e^{-|r_0|}/\left[ e^{r_0}\rho_{11}(0)+e^{-r_0}\rho_{22}(0)\right] ,
    \label{s}
    \end{equation}
and the mean waiting time $T_{\rm undo} = {\tilde P}_1 t_{c,1} + {\tilde
P}_2 t_{c,2}$ until the measurement is undone is \be T_{\rm undo} = T_m  \,
\vert r_0\vert. \label{wait} \ee Several comments are in order about the
main results (\ref{s},\ref{wait}). ($i$) The probability of success $P_s$
given by (\ref{s}) becomes very small for $|r_0|\gg 1$ (when the measurement
result indicates a particular qubit state with good confidence), eventually
becoming $P_s=0$ for a projective measurement, recovering the traditional
statement of irreversibility in this limiting case. ($ii$) In the important
special case when the initial state is pure, the state remains pure during
the entire process. ($iii$) In (\ref{s}), $\rho (0)$ characterizes our
knowledge about an {\it unknown} initial state, in contrast to
(\ref{Bayes-DD}), where $\rho (0)$ represents an ``actual'' state. ($iv$)
Averaging (\ref{s}) over different initial states $\rho^k(0)$ with
corresponding probabilities ${\cal P}_k$ leads to the same result (\ref{s}),
just with $\rho (0)$ replaced by the averaged density matrix $\sum_k {\cal
P}_k\rho^k(0)$.
%(notice that in the
%derivation of this result the probabilities ${\cal P}_k$ should be updated
%using measurement result $r_0$ via the classical Bayes rule).
($v$) If the qubit is entangled with other qubits, the QUD measurement restores
the initial entangled state; the density matrix in (\ref{s}) in this case can be
obtained by tracing over all other qubits. ($vi$) After the initial
measurement, the state evolution is given as a one-to-one map, whose inverse is
known. However, the nonunitarity of the inverse makes its realization impossible
via Hamiltonian evolution; therefore a QUD measurement must be
probabilistic. ($vii$) The undoing probability averaged over the result $r_0$,
$P_{av} = 1 - \mbox{erf}[\sqrt{t/(2T_m)}]$, depends on the ``strength''
$t/T_m$ of the first measurement, but not on the initial state.
%The typical use of Eq.\
%(\ref{Bayes-DD}) is to describe the evolution of a pure state (though it can be
%applied to a mixed state as well), while in Eq.\ (\ref{s}) $\rho (0)$ is
%typically a highly mixed state: for example, if we do not have any initial
%knowledge, we should describe initial state by the completely mixed density
%matrix, $\rho_{11}=\rho_{22}=1/2$, $\rho_{12}=0$, but this depends on how biased
%we are toward pure states \cite{srednicki}.
%    In the case when the measured qubit is entangled with other qubits, the
%density matrix in Eq.\ (\ref{s}) can obviously be obtained by tracing over
%all other qubits.
%    It is important to mention
%that
%    It is easy to show that Eq.\ (\ref{s}) can be rewritten as
%$s=e^{-|r_0|}(\rho^m_{11} e^{-r_0}+\rho^m_{22}e^{r_0})$ in terms of the density
%matrix $\rho^m$ after the first measurement.

{\it Phase qubit.}--- The second explicit example of erasing information and
undoing a quantum measurement is for a superconducting ``phase'' qubit
\cite{Martinis,Katz}. The system (similar to the ``flux'' qubit) is comprised
of a superconducting loop interrupted by one Josephson junction [Fig.\
2(a)], which is controlled by an external flux $\phi_e$. Qubit states
$|1\rangle$ and $|2\rangle$ [Fig.\ 2(b)] correspond to the two lowest states in
a quantum well with potential energy $V(\phi )$, where $\phi$ is the
superconducting phase difference across the junction (for consistency with
the previous example, we do not use the more traditional notation $|0\rangle$
and $|1\rangle$). The qubit is measured by lowering the barrier (which is
controlled by $\phi_e$), so that the upper state $|2\rangle$ tunnels into
the continuum with rate $\Gamma$, while state $|1\rangle$ does not
tunnel out. The tunneling event is sensed by a two-junction detector SQUID
inductively coupled to the qubit [Fig.\ 2(a)].

\begin{figure}[tb]
\includegraphics[width=6.5cm]{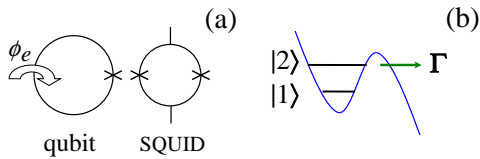}
%\vspace{0.3cm}
\caption{(a) Schematic of a phase qubit controlled by an external flux
$\phi_e$ and inductively coupled to the detector SQUID. (b) Energy profile
$V(\phi )$ with quantized levels representing the qubit states. The
tunneling event is sensed by the SQUID.} \label{fig2}\end{figure}

    For sufficiently long tunneling time $t$, $\Gamma t \gg 1$, the measurement
corresponds to the usual collapse:  the qubit state is either projected onto the lower
state $|1\rangle$ (if no tunneling is recorded) or destroyed (if tunneling
happens). However, if the barrier is raised after a finite time $t \sim
\Gamma^{-1}$, the measurement is weak: the qubit state
is still destroyed if tunneling happens, while in the case of no tunneling
(a null-result measurement) the qubit density matrix evolves in the rotating
frame as \cite{Dalibard,Katz}
    \begin{equation}
\frac{\rho_{11}(t)}{\rho_{22}(t)}= \frac{\rho_{11}(0)}{\rho_{22}(0)\,
e^{-\Gamma t}} , \,\,\, \frac{\rho_{12}(t)}{\sqrt{\rho_{11}(t)\rho_{22}(t)}}=
\frac{\rho_{12}(0)\, e^{-i\varphi (t)}}{\sqrt{\rho_{11}(0)\rho_{22}(0)}},
    \label{Bayes-phase}
    \end{equation}
where the phase $\varphi (t)$ accumulates because of the change of energy
difference between states $|1\rangle$ and $|2\rangle$ when the barrier
is lowered by changing $\phi_e$. Notice that except for the effect of the extra
phase $\varphi (t)$, the qubit evolution (\ref{Bayes-phase}) is similar to
the qubit evolution in the previous example; in particular, it also
represents an ideal measurement which does not decohere the qubit, and
has a clear Bayesian interpretation. Formally, the evolution
(\ref{Bayes-phase}) corresponds to the measurement result $r=\Gamma t/2$ in
Eq.\ (\ref{Bayes-DD}). The coherent non-unitary evolution
(\ref{Bayes-phase}) has been experimentally verified in Ref.\ \cite{Katz}
using tomography of the post-measurement state (in \cite{Katz} the product
$\Gamma t$ was actually varied by changing the tunneling rate $\Gamma$, while
keeping the duration $t$ constant).

{\it Measurement undoing for the phase qubit.}---
A slight modification of the experiment \cite{Katz} can be used to
demonstrate measurement undoing.  Suppose the tunneling event did not happen
during the first weak measurement, so the evolution (\ref{Bayes-phase}) has
occurred. The undoing of this measurement consists of three steps:
$(i)$ Exchange the amplitudes of states $|1\rangle$ and
$|2\rangle$ by the application of a $\pi$-pulse, $(ii)$ perform another weak
measurement, identical to the first measurement, $(iii)$ apply
a second $\pi$-pulse. If the tunneling event did not occur during the second
measurement, then the information about the initial qubit state is erased (both
basis states have equal likelihood for two null-result measurements).
Correspondingly, according to Eq.\ (\ref{Bayes-phase}) (which is applied for
the second time with exchanged indices $1\leftrightarrow 2$), any initial
qubit state is fully restored (notice that the phase $\varphi$ is also
canceled).

    The success probability $P_s$ for the undoing procedure is just the
probability that the tunneling does not happen during the second
measurement. If we start with the qubit state $\rho (0)$, the state after the
first measurement is given by Eq.\ (\ref{Bayes-phase}). After the
$\pi$-pulse, the occupation of the upper state is $\rho_{22}'=
\rho_{11}(0)/[\rho_{11}(0)+\rho_{22}(0)\,e^{-\Gamma t}]$, so the success
probability  $P_s=1-\rho_{22}' (1-e^{-\Gamma t})$ can be expressed as
    \begin{equation}
P_s= e^{-\Gamma t}/\left[ \rho_{11}(0)+e^{-\Gamma t}\rho_{22}(0)\right],
    \label{s-phase}
    \end{equation}
which formally coincides with Eq.\ (\ref{s}) for $r=\Gamma t/2$.
    While measurement undoing is most important for an unknown state, in the
demonstration experiment the initial state can be known, and
tomography of the final state can be used to check that it is identical to
the initial state.
%By varying the initial state, it is possible to show that any
%state is restored by measurement undoing.

{\it General theory of measurement undoing.}---
   Applying POVM formalism \cite{Nielsen}, we can describe a general
quantum measurement with result $r$ by a linear operator $M_r$, so that for
an initial state $\rho$ the probability of result $r$ is $P_r(\rho)
=\mbox{Tr}(E_r \rho)$, where $E_r=M_r^\dagger M_r$, and the state after
measurement is $\tilde{\rho}=M_r \rho M_r^\dagger /\mbox{Tr}(E_r \rho)$. Here
$E_r$ is a positive Hermitian operator, obeying the completeness relation
$\sum_r E_r =1$. In order to undo this measurement, we should apply the
inverse operation characterized by $L_r=C M_r^{-1}$, where $C$ is a complex
number. Such an operation is physical ({\it i.e.} can be realized by a
second measurement yielding a ``lucky'' result) only if all eigenvalues of
$L_r^\dagger L_r$ are not larger than 1 (otherwise completeness cannot be
satisfied), which leads to the upper bound $|C|^2\leq \min_i p_i$, where
$\{p_i\}$ is the set of eigenvalues of $E_r$. Therefore, the probability
$P_s=\mbox{Tr}(L_r^\dagger L_r\tilde{\rho})$ of the lucky result
corresponding to $L_r$ is bounded by $(\min_i p_i)/P_r(\rho )$. Finally
recalling that $\{p_i\}$ are probabilities of the result $r$ for
eigenvectors of $E_r$, we find the upper bound for the probability of
successful undoing (similar to the result of \cite{Ueda-99}):
    \be
P_s \le (\min P_r)/P_r(\rho),
    \label{s-gen}
    \end{equation}
where $\rho$ is the initial state and $\min P_r$ is the probability
of the result $r$ minimized over all possible initial states. Notice
that averaging of $P_s$ over the result $r$ makes it independent of
the initial state: $P_{av}=\sum_r (\min P_r)$.

    Let us compare the general upper bound (\ref{s-gen}) for $P_s$
 with the results (\ref{s}) and (\ref{s-phase}) of the two previous
 examples.
For the QPC measurement of the DQD qubit described by Eq.\
(\ref{Bayes-DD}), the operator $E_r$ is diagonal in the measurement
basis $|1\rangle$, $|2\rangle$ and has matrix elements $p_i =(\pi
S_I/t)^{-1/2} \exp [-({\bar{I}-I_i})^2 t/S_I] \, d{\bar I}$,
$i=1,2$, related to the probability densities of the continuous
variable $\bar I$. Then the upper bound (\ref{s-gen}) becomes $\min
(p_1,p_2)/(p_1\rho_{11}+p_2\rho_{22})$, which coincides with Eq.\
(\ref{s}) because $p_1/p_2=e^{2r}$. We conclude that our undoing
strategy is optimal, since the upper bound (\ref{s-gen}) is reached.
    For the example of phase qubit measurement, the operator $E_r$
corresponding to the null-result measurement (no tunneling) is also
diagonal in the measurement basis $|1\rangle$, $|2\rangle$, and has
matrix elements 1 and $e^{-\Gamma t}$ respectively. Again, $P_s$
given by (\ref{s-phase}) reaches the upper bound (\ref{s-gen}), thus
confirming the optimality of the analyzed undoing procedure.
%{\bf note the phase qubit has a phase, not included in
%the pure POVM case.}

{\it Explicit general procedure of measurement undoing.---} We briefly
discuss a procedure to undo (in principle) an arbitrary one-to-one
measurement $M_r$ for any number $N$ of entangled charge qubits, using
unitary rotations and measurement by a QPC with an extremely strong
nonlinearity, so that tunneling (with rate $\gamma$) occurs in the QPC only
when all qubits are in the state $|1\rangle$. For simplicity assume
$M_r=\sqrt{E_r}$ (the generalization is trivial). In the basis of $2^N$
vectors ${|i\rangle}$ diagonalizing $E_r$, the desired undoing operator
$L_r=\sqrt{\min_j p_j}\, M_r^{-1}$ (see above) is also diagonal:
$L_{r,ii}=\sqrt{(\min_j p_j)/p_i}$. It can be realized in $2^N$ steps. The
$i$th step consists of a unitary rotation of the vector $|i\rangle$ into the
state $|11\dots 1\rangle$, measurement by the QPC for duration
$\tau_i=-\gamma^{-1} \ln L_{r,ii}^2$, and the reverse unitary rotation. In
each step the no-tunneling result corresponds to the measurement operator,
which is almost unity, except for diagonal matrix element $e^{-\gamma
\tau_i/2}=L_{r,ii}$ for the vector $|i\rangle$. The measurement undoing
procedure is successful if no tunneling occurred in all steps. The
corresponding success probability reaches the upper bound (\ref{s-gen}).

{\it Undoing continuous measurement of an evolving charge qubit.}--- In our
first example we have assumed for simplicity that the qubit is not
undergoing Hamiltonian evolution during the measurement process to be
undone. We briefly note that a QUD measurement is also possible when Hamiltonian
evolution is included. In this case the qubit evolution is described by the
Bayesian equations \cite{Kor-99}, which can be used to find the operator
$M_r$ from the measurement record $I(t)$ (the non-normalized version of the
Bayesian equations is more appropriate for this purpose). Then the desired
undoing operator $L_r$ can be realized in three steps, corresponding to the
singular value decomposition of $L_r$: unitary rotation, continuous
measurement by QPC (with internal qubit dynamics turned off), and one more
unitary rotation. (The details of this calculation and the explicit undoing
procedure will be presented elsewhere.)

The authors thank A. Kitaev for useful discussions and J. Dowling for
suggesting the term quantum un-demolition measurement.  This work was
supported by NSA and DTO under ARO grant W911NF-04-1-0204 (ANK), and AFRL
Grant F30602-01-1-0594, AFOSR Grant FA9550-04-1-0206, and TITF Grant
2001-055 (ANJ).

\vspace{-0.4cm}

%\begin{references}


\begin{thebibliography}{99}
\bibitem{S-normalization} We use normalization of the shot noise, in which
    $S_{I}=2eI(1-{\cal T})$, where ${\cal T}$ is the QPC transparency.
\end{thebibliography}
\end{document}